\documentclass[prd,twocolumn,showpacs,amsmath,amssymb]{revtex4}
\usepackage{psfig}% Include figure files
\usepackage{dcolumn}% Align table columns on decimal point
\usepackage{bm}% bold math
\newcommand{\kskl}{K_S^0 K_L^0}
\newcommand{\ks}{K_S^0}
\newcommand{\kl}{K_L^0}
\newcommand{\BR}{{\cal B}}
\newcommand{\eff}{\varepsilon}
\newcommand{\pspp}{\psi''}
\newcommand{\psp}{\psi'}
\newcommand{\jpsi}{J/\psi}

\newcommand{\ccbar}{c\bar{c}}
\newcommand{\EE}{e^+e^-}
\newcommand{\MM}{\mu^+\mu^-}

\newcommand{\pp}{\pi^+\pi^-}

\newcommand{\ppb}{p\overline{p}}

\newcommand{\kskn}{K^{*0}(892)\overline{K^0}+c.c.}

\newcommand{\jpsipp}{\pi^+\pi^-J/\psi}

\newcommand{\rhopi}{\rho\pi}
\newcommand{\ra}{\rightarrow}
\newcommand{\jpsito}{J/\psi \rightarrow }

\newcommand{\pspto}{\psi' \rightarrow }
\newcommand{\psppto}{\psi'' \rightarrow }
\newcommand{\beq}{\begin{equation}}
\newcommand{\eeq}{\end{equation}}
\newcommand{\beqn}{\begin{eqnarray}}
\newcommand{\eeqn}{\end{eqnarray}}
\newcommand{\beqns}{\begin{eqnarray*}}
\newcommand{\eeqns}{\end{eqnarray*}}
\newcommand{\bfg}{\begin{figure}}
\newcommand{\efg}{\end{figure}}
\newcommand{\bitm}{\begin{itemize}}
\newcommand{\eitm}{\end{itemize}}
\newcommand{\bnum}{\begin{enumerate}}
\newcommand{\enum}{\end{enumerate}}
\newcommand{\btbl}{\begin{table}}
\newcommand{\etbl}{\end{table}}
\newcommand{\btbu}{\begin{tabular}}
\newcommand{\etbu}{\end{tabular}}
\begin{document}
\normalsize
\parskip=5pt plus 1pt minus 1pt

\title{Search for $\kskl$ in $\pspp$ Decays}
\author{
M.~Ablikim$^{1}$, J.~Z.~Bai$^{1}$, Y.~Ban$^{10}$,
J.~G.~Bian$^{1}$, X.~Cai$^{1}$, J.~F.~Chang$^{1}$,
H.~F.~Chen$^{15}$, H.~S.~Chen$^{1}$, H.~X.~Chen$^{1}$,
J.~C.~Chen$^{1}$
% ...
\\(BES Collaboration)\\
} \affiliation{
$^1$ Institute of High Energy Physics, Beijing 100039, People's Republic of China\\
$^2$ China Center for Advanced Science and Technology(CCAST),
Beijing 100080,
People's Republic of China\\
$^3$ Guangxi Normal University, Guilin 541004, People's Republic of China\\
$^4$ Henan Normal University, Xinxiang 453002, People's Republic of China\\
$^5$ Huazhong Normal University, Wuhan 430079, People's Republic of China\\
$^6$ Hunan University, Changsha 410082, People's Republic of China\\
$^7$ Liaoning University, Shenyang 110036, People's Republic of China\\
$^8$ Nanjing Normal University, Nanjing 210097, People's Republic of China\\
$^9$ Nankai University, Tianjin 300071, People's Republic of China\\
$^{10}$ Peking University, Beijing 100871, People's Republic of China\\
$^{11}$ Shandong University, Jinan 250100, People's Republic of China\\
$^{12}$ Shanghai Jiaotong University, Shanghai 200030, People's Republic of China\\
$^{13}$ Sichuan University, Chengdu 610064, People's Republic of China\\
$^{14}$ Tsinghua University, Beijing 100084, People's Republic of China\\
$^{15}$ University of Science and Technology of China, Hefei 230026, People's Republic of China\\
$^{16}$ Wuhan University, Wuhan 430072, People's Republic of China\\
$^{17}$ Zhejiang University, Hangzhou 310028, People's Republic of China\\
$^{\ast}$ Current address:  University of Michigan, Ann Arbor, MI 48109 USA \\
$^{\dagger}$ Current address: Purdue University, West Lafayette,
Indiana 47907, USA. }
\date{\today}

\begin{abstract}

    $\kskl$ from $\pspp$ decays is searched for using
the $\pspp$ data collected by BESII at BEPC, the upper limit of
the branching fraction is determined to be \( \BR(\psppto \kskl) <
2.1\times 10^{-4}\) at 90\% C.~L. The measurement is compared with
the prediction of the $S$- and $D$-wave mixing model of the
charmonia, based on the measurements of the branching fractions of
$\jpsito \kskl$ and $\pspto \kskl$.

\end{abstract}


\maketitle

\section{Introduction}

From the perturbative QCD (pQCD), it is expected that both $\jpsi$
and $\psp$ decaying into light hadrons are dominated by the
annihilation of $\ccbar$ into three gluons or a virtual photon,
with widths proportional to the square of the wave function at the
origin~\cite{appelquist}. This yields the pQCD ``12\% rule'', that
is
\[
Q_h =\frac{{\cal B}_{\psp \ra h}}{{\cal B}_{\jpsi \ra h}}
=\frac{{\cal B}_{\psp \ra \EE}}{{\cal B}_{\jpsi \ra \EE}} \approx
 12\%.
\]

Following the first observation of its violation in $\rhopi$ and
$K^{*+}K^-+c.c.$ modes by Mark II~\cite{mk2}, BES has measured
many two-body modes of $\psp$ decays, among which some obey the
12\% rule while others violate it~\cite{besres}. There have been
many theoretical efforts trying to solve the
puzzle~\cite{puzzletheory}, however, none explains all the
existing experimental data satisfactorily and naturally.

A most recent explanation of the ``$\rhopi$ puzzle'' using the
$S$- and $D$-wave charmonia mixing was proposed by
Rosner~\cite{rosnersd}. In this scheme, the mixing of  $\psi(2^3
S_1)$ state and $\psi(1^3 D_1)$ is in such a way which leads to
almost complete cancellation of the decay amplitude of $\psp
\rightarrow \rhopi$, and enhanced decay rate of $\pspp$. A study
on the measurement of $\pspp \ra \rhopi$ in $\EE$ experiments
shows that with the decay rate predicted by the $S$- and $D$-wave
mixing, the interference between the three-gluon decay amplitude
of the $\pspp$ and the continuum one-photon amplitude is
destructive so the observed cross section is very
small~\cite{wympspp}, which is in agreement with the upper limit
of the $\rhopi$ cross section at the $\pspp$ peak by Mark
III~\cite{mk3}. Although this need to be further tested by high
luminosity experiment operating at the $\pspp$ mass energy, such
as CLEOc~\cite{cleoc}, it already implied that ${\cal B}(\pspp
\rightarrow \rhopi)$ is most probably at the order of $10^{-4}$,
in agreement with the prediction of the $S$- and $D$-wave mixing
scheme.

If the $S$- and $D$-wave mixing is the key for solving the
$\rhopi$ puzzle, it applies to other decay modes as well, such as
pseudoscalar pseudoscalar (PP) mode like $\kskl$. Recently, BES
collaboration reported the branching fractions of $\kskl$ final
state in $\jpsi$ and $\psp$ decays~\cite{besksklj,beskskl}:
 \beqn
 \BR(\jpsi \ra\kskl)=(1.82\pm0.04\pm0.13)\times10^{-4}, \nonumber \\
 \BR(\psp \ra \kskl)=(5.24\pm0.47\pm0.48)\times10^{-5}. \nonumber
 \eeqn
These results yield $Q_{\kskl} =(28.8 \pm 3.7)\%$, which is
enhanced relative to the 12\% rule by more than 4$\sigma$.

By assuming that the pQCD 12\% rule holds for $\kskl$ mode between
$\jpsi$ and $\psi(2^3S_1)$, in the $S$- and $D$-wave charmonia
mixing scheme, Ref.~\cite{wmykskl} predicts the decay rate of
$\pspp \ra \kskl$ in a range, that is
 \[ 0.12\pm 0.07 \le 10^{5}\times \BR(\pspp \ra \kskl) \le 3.8\pm 1.1. \]
Here the upper bound corresponds to $\phi= 0^\circ$ and the lower
bound to $\phi= 180^\circ$, where $\phi$ is the relative phase
between $\langle \kskl|1^3D_1 \rangle$ and $\langle \kskl|2^3S_1
\rangle$. The uncertainties are due to the mixing angle $\theta$
between $\psi (2^3 S_1)$ and $\psi (1^3 D_1)$ states, and the
measurements of $\BR(\psp \ra \kskl)$ and $\BR(\jpsi \ra \kskl)$.

In this paper, we report a search for $\psppto \kskl$ at BESII.

\section{The Experiment}

The data used for the analysis  are taken with the BESII detector
at the BEPC storage ring in the vicinity of the $\pspp$ peak ($\pm
4$~MeV around $\pspp$ nominal mass). The data sample corresponds
to a total of $17.7(1\pm 5\%)~pb^{-1}$ luminosity as determined
from large angle Bhabha events~\cite{0307028}.

The BES is a conventional solenoidal magnet detector that is
described in detail in Ref.~\cite{bes}, BESII is the upgraded
version of the BES detector~\cite{bes2}. A 12-layer vertex chamber
(VC) surrounding the beam pipe provides trigger information. A
forty-layer main drift chamber (MDC), located radially outside the
VC, provides trajectory and energy loss ($dE/dx$) information for
charged tracks over $85\%$ of the total solid angle.  The momentum
resolution is $\sigma _p/p = 0.017 \sqrt{1+p^2}$ ($p$ in
$\hbox{\rm GeV}/c$), and the $dE/dx$ resolution for hadron tracks
is $\sim 8\%$. An array of 48 scintillation counters surrounding
the MDC  measures the time-of-flight (TOF) of charged tracks with
a resolution of $\sim 200$ ps for hadrons.  Radially outside the
TOF system is a 12 radiation length, lead-gas barrel shower
counter (BSC).  This measures the energies of electrons and
photons over $\sim 80\%$ of the total solid angle with an energy
resolution of $\sigma_E/E=22\%/\sqrt{E}$ ($E$ in GeV).  Outside of
the solenoidal coil, which provides a 0.4~Tesla magnetic field
over the tracking volume, is an iron flux return that is
instrumented with three double layers of  counters that identify
muons of momentum greater than 0.5~GeV/$c$.

\section{Monte Carlo}

Monte Carlo (MC) is used for mass resolution and detection
efficiency determination, as well as the background study.

For the signal channel, $\psppto \kskl$, the angular distribution
of $\ks$ or $\kl$ is generated as $\sin^2\theta$, where $\theta$ is
the polar angle in laboratory system. $\kl$ is allowed to decay
according to its lifetime in the detector and only $\ks \ra \pp$
is generated. For this study, 10000 events are generated.

One of the main backgrounds is from $\EE \ra \gamma \gamma
(\gamma)$ events, with one photon converted into $\EE$ pair in the
detector material. This is studied with a MC sample which is 4
times as large as in the real data.

Another background channel is $\kskn$ from $\pspp$ decays or $\EE$
direct production. 10000 events are generated for studying this
background, which is about 100 times more than in the real data.

2M $D\bar{D}$ pairs are generated for studying the background
channels with c-quark production. This sample is about 14 times
more than the real data sample.

The continuum channels from u, d, and s quark fragmentation are
generated with JETSET7.4~\cite{jetset}, the MC sample is about 4
times as large as in the real data.

The simulation of the detector response is a Geant3 based package,
where the interactions of the secondary particles with the
detector material are simulated. Reasonable agreement between data
and Monte Carlo simulation has been observed in various testing
channels including $\EE \ra \EE$, $\EE\ra \MM$, $\jpsito \ppb$ and
$\pspto \jpsipp, \jpsito \ell^+\ell^-$ $(\ell=e,\mu)$.

\section{Event selection}

The event selection criteria are all used in the analyses of the
same final states at $\jpsi$~\cite{besksklj} and $\psp$
energy~\cite{beskskl}. They are listed here for a easy reference.

\begin{enumerate}
\item   The number of charged tracks is required to be two
        with net charge zero. Each track should have good helix
        fit so that the error matrix of the track fitting is
        available for secondary vertex finding. The track is
        required to be within $|\cos\theta|<0.80$, where
        $\theta$ is the polar angle of the track in MDC in the
        laboratory system.
\item   The two tracks are assumed to be $\pi^+$ and $\pi^-$,
        to find the intersect of the two tracks near the interaction
        point, which will be taken as the secondary vertex.
        The $\pi^+\pi^-$ mass is required to be within 2$\sigma$
        of the MC predicted mass resolution (8.9~MeV/$c$). The decay
        length in $xy$-plane, $L_{xy}>0.01$~m is required.
\item   The sum of the total energy of the photon candidates
        $E_{\gamma}^{tot}<1.0$~GeV is used to remove the
        $\EE\ra \gamma\gamma (\gamma)$ backgrounds, with one photon converted
        into $\EE$ pairs in the detector material. A neutral cluster is
        considered to be a photon candidate when the
        angle between the nearest charged track and the cluster in the
        $xy$ plane is greater than $15^{\circ}$, the first hit is in the
        beginning 6 radiation lengths, and the angle between the cluster
        development direction in the BSC and the photon emission direction
        in $xy$ plane is less than $37^{\circ}$.
\end{enumerate}

The above selection criteria are exactly the same as used for
$\pspto \kskl$ analysis~\cite{beskskl}. To be unbiased with the
expected small signal in the analysis, we try to fix the event
selection criteria before looking at real data. By looking at the
$\ks$ momentum distributions of the background MC samples
generated, it is found that the background from the $\EE\ra \gamma
\gamma(\gamma)$ and $\kskn$ is still very large. Then the
selection criteria used for $\jpsito \kskl$
selection~\cite{besksklj} are further applied to reduce the
background level.

\bnum \item[4.]  The total BSC energy associated with the two
charged tracks less than 1.0~GeV or the total $XSE$ (the
difference from the expected $dE/dx$ for the electron hypothesis
divided by the $dE/dx$ resolution) is less than $-4$. \item[5.]
The opening angle between the two charged tracks larger than
$20^\circ$. \item[6.] $E_{\gamma}^{lft}<0.1$~GeV, where
$E_{\gamma}^{lft}$ is the sum of the energies of the photon
candidates outside a cone around the direction of $\kl$
($\cos\theta<0.95$). \enum

Cuts 4 and 5 are used to reduce the gamma conversion background
and cut 6 is used to reduce the $\kskn$ background. After all the
above cuts, there is no events left in the
$\EE\ra\gamma\gamma(\gamma)$ MC sample, and the remaining $\kskn$
events in the signal region is less than 1 after normalized to the
$\kskn$ cross section~\cite{hd03_wangp} and the luminosity of the
data sample.

The signal region is defined as the $\ks$ momentum to be larger
than 1.737~GeV/$c$, which is 2$\sigma$ ($\sigma=42$~MeV/$c$) lower
than the MC predicted $\ks$ momentum for the signal channel.

\begin{figure}[htb]
\centerline{\psfig{file=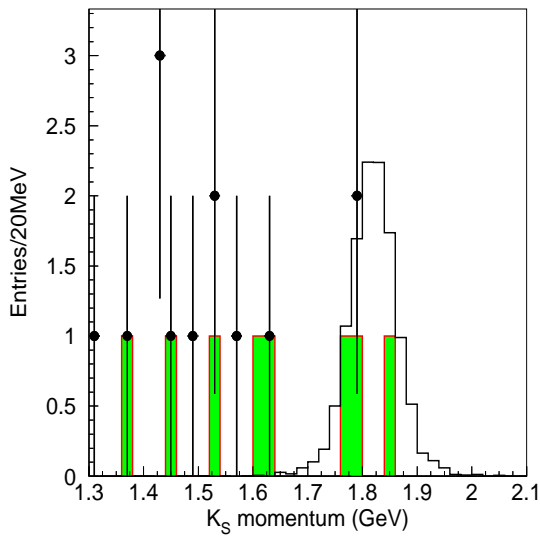,height=7cm,width=7.0cm}}
\caption{The $\ks$ momentum distribution. Data are shown by dots
with error bars, the $\ks$ mass sidebands background is shown by
shaded histogram. The blank histogram is the MC simulated signal
channel events, not normalized.} \label{pksside}
\end{figure}

After requiring all above criteria, the $\ks$ momentum
distribution of real data is shown in Fig.~\ref{pksside} as the
black dots with error bars, the distribution of the events in the
$\ks$ mass sidebands is also shown (shaded histogram), we can see
that there is no clear difference between events in $\ks$ mass
region and those in $\ks$ mass sidebands. In the same plot, we
also give the MC predicted position of the signal events (blank
histogram). It can be seen that there is no clear signal in data
at the signal region. So we conclude that there is no $\kskl$
signal observed, and the upper limit of the $\psppto \kskl$
branching fraction will be determined based on the two candidates
in the signal region.

\section{Efficiencies and systematic errors}

The detection efficiency of the signal is estimated with 10000
Monte Carlo simulated events, one gets $\eff_{MC}=(41.44\pm
0.49)\%$, where the error is due to the statistics of the Monte
Carlo sample. The trigger efficiency of $\kskl$ events, which is
lower because of the $\ks$ decays, is measured to be $(76.0\pm
1.8)\%$~\cite{beskskl}. It is found that the reconstruction
efficiency of the $\ks$ in Monte Carlo is a bit higher than that
in data~\cite{besksklj}, a correction factor of $(96.3\pm 3.3)\%$
should be applied to the Monte Carlo simulation. After taking into
account all these factors, the global efficiency is
$\eff=30.33\%$.

The systematic error for the branching fraction measurement comes
from the efficiencies of photon ID, secondary vertex finding, MDC
tracking, trigger, the branching fraction used, number of $\pspp$
events, $\ks$ mass cut, angular distributions and so on. All these
have been studied extensively in the analyses of $\jpsi$ and
$\pspto \kskl$~\cite{beskskl,besksklj}, and they are borrowed from
the two analyses directly.

Table.~\ref{sys} lists the systematic error from all sources. The
uncertainties of the $\BR(\psppto\EE)$ and luminosity will affect
the determination of the total number of the $\pspp$ events, the
former comes from the PDG~\cite{pdg} while the later is from the
measurement used in Ref.~\cite{0307028}.

\begin{table}[htbp]
\caption{Summary of the systematic errors.}
\begin{center}
\begin{tabular}{r|c}
\hline\hline Source  & Systematic errors (\%) \\\hline
MC statistics               &  1.3              \\
$E_{\gamma}^{tot}$          &  2                \\
$E_{\gamma}^{lft}$          &  1.3                \\
$2^{nd}$ vertex finding     &  3.4  \\
MDC tracking                &  4                \\
Trigger efficiency          &  2.4  \\
$\BR(\psppto\EE)$           &  15               \\
Luminosity                  &   5               \\
$\BR(\ks \ra \pp)$          &  0.4                   \\\hline
Total $\sigma^{sys}$        &  17
\\\hline\hline
\end{tabular}
\end{center}
\label{sys}
\end{table}

Add the errors from all the sources in quadrature, the total
systematic error is 17\%.

\section{Results and discussion}

The upper limit of the branching fraction of $\psppto \kskl$
calculated with \[ \BR(\psppto
\kskl)<\left.\frac{n^{obs}_{UL}/\eff}
     {N_{\pspp}\BR(\ks\ra \pp)(1-\sigma^{sys})}
     \right..
\]
 Where $n^{obs}_{UL}$ is the upper limit of the observed
number of events, which is 5.32 for 2 observed events at 90\%
C.~L. assuming there is no background. $\eff$ is the global
efficiency for the signal channel. $N_{\pspp}$ is the number of
$\pspp$ events, calculated with the total luminosity and the
resonance parameters listed by PDG~\cite{pdg}. The systematic
error of the measurement is considered by introducing
$1-\sigma^{sys}$ in the denominator of the formula for branching
fraction calculation.

Using numbers got above (listed in Table.~\ref{br}), one gets, at
90\% C.~L.,
\[ \BR(\psppto \kskl)  <2.1\times 10^{-4}. \]

\begin{table}[htbp]
\caption{Numbers used in the calculation of upper limit of the
branching fraction.}
\begin{center}
\begin{tabular}{r|c}
\hline\hline quantity    & Value \\\hline
$n^{obs}_{UL}$     & $5.32$ \\
$\eff$          & $30.33\%$ \\
$N_{\pspp}$   &  $1.45\times 10^5$ \\
$\BR(\ks \ra \pp)$  &  $0.6860$ \\
$\sigma^{sys}$      &  $17\%$    \\
$\BR(\psppto \kskl)<$ & $2.1\times 10^{-4}$  \\
\hline\hline
\end{tabular}
\end{center}
\label{br}
\end{table}

Comparing with the corresponding theoretical calculation of the
branching fraction based on the $S$- and $D$-wave mixing model and
the 12\% rule, current upper limit is still well above the upper
bound of the prediction~\cite{wmykskl} of $3.8\times 10^{-5}$. To
further pin down the upper limit of the branching fraction, thus
to check the validity of the ``Rosner's assumption" and the
solution of the $\kskl$ enhancement puzzle observed in $\psp$ and
$\jpsi$ decays, a larger data sample is needed. The existing
$55~pb^{-1}$ $\pspp$ data and the planned $3~fb^{-1}$ $\pspp$ data
samples from CLEOc~\cite{cleoc} will be obviously helpful for this
study.

\section{Summary}

Flavor SU(3) breaking process $\kskl$ is searched for in $\pspp$
decays with BESII data sample at $\pspp$ energy, and the upper
limit of the branching fraction is determined to be \( \BR(\psppto
\kskl) <2.1 \times 10^{-4} \). The upper limit is still above the
upper bound of the prediction~\cite{wmykskl}.

\acknowledgments

The BES collaboration thanks the staff of BEPC for their hard
efforts and the members of IHEP computing center for their helpful
assistance. This work is supported in part by the National Natural
Science Foundation of China under contracts Nos. 19991480,
10225524, 10225525, the Chinese Academy of Sciences under contract
No. KJ 95T-03, the 100 Talents Program of CAS under Contract Nos.
U-11, U-24, U-25, and the Knowledge Innovation Project of CAS
under Contract Nos. U-602, U-34 (IHEP), and by the National
Natural Science Foundation of China under Contract No. 10175060
(USTC), and No. 10225522 (Tsinghua University).

\end{document}